\documentclass{article}

\usepackage{arxiv}

\usepackage[utf8]{inputenc} 
\usepackage[T1]{fontenc}    
\usepackage{hyperref}       
\usepackage{url}            
\usepackage{booktabs}       
\usepackage{amsfonts}       
\usepackage{nicefrac}       
\usepackage{microtype}      
\usepackage{lipsum}		
\usepackage{graphicx}
\usepackage{natbib}
\usepackage{doi}
\usepackage{blkarray}
\usepackage{multirow}
\newtheorem{theorem}{Theorem}
\newtheorem{example}[theorem]{Example}
\usepackage{colortbl}
\usepackage{xcolor}
\definecolor{Gray}{gray}{0.65}
\definecolor{Gray}{gray}{0.25}
\DeclareMathAlphabet{\pazocal}{OMS}{zplm}{m}{n}
\SetMathAlphabet\pazocal{bold}{OMS}{zplm}{bx}{n}

\title{Identifying critical higher-order interactions in complex networks}

\author{
  Mehmet Emin Aktas\\
  Department of Mathematics and Statistics\\
  University of Central Oklahoma\\
  Edmond, OK 73034 \\
  \texttt{maktas@uco.edu} \\
   \And
 Thu Nguyen \\
  Department of Computer Science\\
  University of Central Oklahoma\\
  Edmond, OK 73034 \\
  \texttt{tnguyenthuyanh@uco.edu} \\
  \And
 Sidra Jawaid\\
  Department of Mathematics and Statistics\\
  University of Central Oklahoma\\
  Edmond, OK 73034 \\
  \texttt{sjawaid@uco.edu} \\
  \And
 Rakin Riza \\
  Department of Computer Science\\
  University of Central Oklahoma\\
  Edmond, OK 73034 \\
  \texttt{rriza@uco.edu} \\
  \And
 Esra Akbas \\
  Department of Computer Science\\
  Oklahoma State University\\
  Stillwater, OK 74078 \\
  \texttt{eakbas@okstate.edu} \\
}

\renewcommand{\shorttitle}{Identifying critical higher-order interactions in complex networks}

\hypersetup{
pdftitle={Identifying critical higher-order interactions in complex networks},
pdfsubject={q-bio.NC, q-bio.QM},
pdfauthor={Mehmet Emin Aktas},
pdfkeywords={Hypergraphs, influential higher-order interactions, information diffusion, simplical Laplacian},
}

\begin{document}
\maketitle

\begin{abstract}
Information diffusion on networks is an important concept in network science observed in many situations such as information spreading and rumor controlling in social networks, disease contagion between individuals, cascading failures in power grids. The critical interactions in networks are the ones that play critical roles in information diffusion and primarily affect network structure and functions. Besides, interactions can occur between not only two nodes as pairwise interactions, i.e., edges, but also three or more nodes, described as higher-order interactions. This report presents a novel method to identify critical higher-order interactions. We propose two new Laplacians that allow redefining classical graph centrality measures for higher-order interactions. We then compare the redefined centrality measures using the Susceptible-Infected-Recovered (SIR) simulation model. Experimental results suggest that the proposed method is promising in identifying critical higher-order interactions.
\end{abstract}



\section{Introduction}
Information diffusion on networks is an important concept in network science observed in many situations such as information spreading and rumor controlling in social networks, disease contagion between individuals, cascading failures in power grids. During the diffusion, some network structures play critical roles and primarily affect network structure and functions. For example, in social networks, one can spread messages in the network quickly through the critical nodes~\cite{kempe2003maximizing}; in epidemic networks, one may reduce the diffusion of epidemic with controlling influential nodes~\cite{wang2020preventing}. Therefore, identifying critical (influential) nodes and edges has practical importance in network science. There are many studies in the literature for the critical node detection problem in networks. Typical methods are based on network topology. Some studies use nodes' degree, such as degree centrality~\cite{bonacich1972factoring} and H-index~\cite{lu2016h}, some use paths in networks, such as closeness centrality~\cite{freeman1978centrality} and betweenness centrality~\cite{freeman1977set}, and some use eigendecomposition of graphs, such as PageRank \cite{brin1998anatomy} and DFF centrality \cite{aktas2021influential}. In addition, some researchers use node deletion or contraction to distinguish the importance of nodes~\cite{arulselvan2009detecting,yang2019identification}. 

Besides, critical edges in networks also play a critical role in information diffusion\cite{giuraniuc2005trading,girvan2002community, wang2017identification,zio2012identifying,saito2016detecting,wong2017finding,hamers1989similarity,yu2018identifying,cheng2010bridgeness}. For example, critical edges give important transmission lines in complex power networks. As another example, the identification of critical edges can be helpful to analyze the vulnerability in electrical transmission networks. Recently, many researchers focus on finding critical edges based on network topology. For example, the authors in \cite{giuraniuc2005trading} use the degree of the two nodes that an edge connects to measure the importance of the edge. In \cite{girvan2002community, wang2017identification,zio2012identifying}, the authors use the betweenness centrality of edges to detect critical edges. In other words, they assume that edges connecting two connected components are important. There are also other studies that use flow/reachability\cite{saito2016detecting,wong2017finding}, bridgeness \cite{cheng2010bridgeness}, neighbors \cite{hamers1989similarity}, and clique degrees \cite{yu2018identifying} to measure the edge importance. 

On the other hand, as we see in different real-world applications, such as in human communication, chemical reactions, and ecological systems, interactions can occur between not only two nodes as pairwise interactions, i.e., edges, but also three or more nodes~\cite{battiston2020networks} described as \textit{higher-order} interactions. Hypergraphs can be used to model higher-order interactions in complex systems where entities are represented as nodes, and higher-order interactions among them are represented as hyperedges. For example, in coauthorship networks, nodes represent authors, and hyperedges represent coauthorship between authors. As another example, in drug-drug interaction networks, nodes are substances that make up the drug, and hyperedges correspond to drugs. Besides, critical hyperedges also play a critical role in information diffusion. For example, to find the most influential article in a coauthorship network, we need to find the most influential hyperedge. As another example, in a social network, manufacturers intend to detect influential hyperedges for promoting their products to maximize the number of influenced customers. 

There are a few studies that explore the critical hyperedges in hypergraphs. In \cite{veremyev2017finding,veremyev2019finding,vogiatzis2015integer}, the authors use the betweenness centrality and cliques to find the influential higher-order interactions with a fixed size, however they are only able to compare interactions of the same size. In \cite{nasirian2020detecting,zhao2017efficient}, the authors find influential higher-order interactions based on closeness centrality and are unable to apply other centrality measures. 

In this paper, to address these limitations, we propose two new hypergraph Laplacians based on the diffusion framework that allows us to find the influential higher-order interactions in a hypergraph of any size and with any desired classical centrality measure; one is based on diffusion between fixed size hyperedges, and the other is based on diffusion between all hyperedges. The previously developed hypergraph Laplacians are only defined for special hypergraphs, and more importantly, neglect the relations between hyperedges. Thanks to the proposed Laplacians, we can model the complete relations between hyperedges of any size. Next, we redefine four graph centrality measures, namely DFF, degree, betweenness, and closeness, to hypergraphs and rank higher-order interactions based on these measures. One can also similarly redefine other classical centrality measures, but we believe working with four centralities would be enough to show the effectiveness of the proposed method. For evaluation, we experiment on several undirected real-world network datasets and evaluate the performance using the Susceptible-Infected-Recovered (SIR) simulation model\cite{newman2002spread}. The experimental results suggest that our methods are quite promising in finding influential higher-order interactions.

\section{Methods}

In this section, we start with defining the graph Laplacian. We then present the two hypergraph Laplacians that allow detecting the influential higher-order interactions. Lastly, we present the redefined graph centrality measures. We conclude this section with an illustrative example.    

Let $G$ be a weighted undirected graph. We define the graph \textit{Laplacian} $L$ as $L=D-A$, where $D$ is the weighted degree matrix and $A$ is the weighted adjacency matrix. The graph Laplacian only uses pairwise interactions, i.e., edges, between vertices and ignores higher-order interactions. Furthermore, it only allows to model diffusion between vertices, not higher-order structures. 

To address these concerns, we first represent a complex network with a \textit{hypergraph}. A \textit{hypergraph} $H$ denoted by $H=(V,E=(e_i)_{i \in I})$ on the finite vertex set $V$ is a family $(e_i)_{i \in I}$ ($I$ is a finite set of indexes) of subsets of $V$ called \textit{hyperedges}. In a hypergraph, nodes represent entities and hyperedges represent higher-order interactions in the network. The \textit{size} of a hyperedge is the number of the nodes in the corresponding higher-order interaction. 

We use the diffusion framework on hypergraphs, i.e., the patterns of who is connected to whom, for identifying critical higher-order interactions. Similar to this study \cite{horak2013spectra}, one can define the simplicial Laplacian to model diffusion over a hypergraph with a simplicial complex structure, i.e., subsets of hyperedges are also hyperedges. Let $D_p$ be the \textit{incidence matrix} that encodes which $p$-simplices are incident to which $(p+1)$-simplices. It is defined as
$$
D_p(i,j)= \left\{\begin{tabular}{ll}
 $1$ & if $\sigma_j^p$ is on the boundary of $\sigma_i^{p+1}$ \\
$0$  & otherwise
\end{tabular}\right.
$$
Let $W_p$ be the diagonal weight matrix of the simplices. Then, the $i$-dimensional up Laplacian can be expressed as the matrix
$$\pazocal{L}_i^{up}=W_i^{-1}D_i^{T}W_{i+1}D_i.$$
Similarly, the $i$-dimensional down Laplacian can be expressed as the matrix 
$$ \pazocal{L}_i^{down}=D_{i-1}W_{i-1}^{-1}D_{i-1}^{T}W_i.$$ Lastly, the $i$-dimensional Laplacian in both directions is 
$$~ \pazocal{L}_i^{both}=\pazocal{L}_i^{up}+\pazocal{L}_i^{down}.$$

However, these Laplacians have three critical issues. First, it is defined only for hypergraphs with the simplicial complex structure, which is often not the case in real-world hypergraphs. For example, in a coauthorship network, there may be an article with three coauthors (i.e., a triangle) where each pair of the authors may not have coauthored an article (i.e., edges). Second, for a hyperedge of size $k$, the simplicial Laplacian models the diffusion only \textit{through} the hyperedges of sizes $k-1$ and/or $k+1$. However, in the diffusion framework, information on a hyperedge can diffuse through other hyperedges regardless of their sizes. For example, these Laplacians assume that information from a paper of three coauthors can only diffuse through papers of two and four coauthors, which is not realistic. Third, when we use the simplicial Laplacians in modeling diffusion, we need to assume that information only diffuses \textit{between} fixed size hyperedges. However, a hyperedge can affect other hyperedges regardless of their sizes. For example, based on these Laplacians, information can only diffuse between papers of three coauthors, which is not realistic again.

To address these issues, we develop two hypergraph Laplacians; one is based on diffusion between fixed size hyperedges, $\pazocal{L}_{k}$, and the other is based on diffusion between hyperedges of any size in the hypergraph $H$, $\pazocal{L}_{H}$~\cite{aktas2021hypergraph}. In the simplicial Laplacian, a hyperedge of size $k+1$ is called a $k$-simplex. For example, vertices are called 0-simplices, edges are called 1-simplices and triangles are called 2-simplices. To be consistent with the simplicial Laplacian definition, we prefer to call a hyperedge of size $k+1$ as $k$-simplex while defining our Laplacians. We first update the incidence matrix for simplices of any dimension as follows.
$$
D_{p,r}(i,j)= \left\{\begin{tabular}{ll}
 $1$ & if $\sigma_j^p$ is on the boundary of $\sigma_i^r$ \\
$0$  & otherwise
\end{tabular}\right.
$$
with $\sigma^p$ being a $p$-simplex. Here, $D_{p,r}$ encodes which $p$-simplices are incident to which $r$-simplices. Next, for a hypergraph with the maximum simplex dimension of $n$ (i.e., hyperedge size of $n+1$), Laplacian between $k$-simplices through other simplices is defined as $$
\pazocal{L}_{k}=\pazocal{L}_{k,0}+\pazocal{L}_{k,1}+\cdots+\pazocal{L}_{k,n-1}+ \pazocal{L}_{k,n}
$$
for $k\in \{0,\dots,n\}$, where $$
\pazocal{L}_{k,l}=\left\{\begin{tabular}{ll}
 $W_l^{-1}D_{k,l}^{T}W_{k}D_{k,l}$ & if $k\leq l$ \\
$D_{l,k}W_{k}^{-1}D_{l,k}^{T}W_l$  & if $k>l$ 
\end{tabular}\right.
$$ with $W$ being the diagonal simplex weight matrix. Here, $\pazocal{L}_k$ encodes how $k$-simplices are related to each other where the relations can be through the shared neighboring simplices of any dimension. Next, we improve this Laplacian by considering relations between simplices of any dimension. Using the Laplacian $\pazocal{L}_k$, we define the generalized hypergraph Laplacian, $\pazocal{L}_H$, as the following block matrix
$$
\pazocal{L}_H=\left(
\begin{array}{c|c|c|c|c}
\pazocal{L}_0 & \pazocal{D}_{0,1}^T & \pazocal{D}_{0,2}^T & \cdots & \pazocal{D}_{0,n}^T \\ \hline
\pazocal{D}_{0,1} & \pazocal{L}_1 & \pazocal{D}_{1,2}^T & \cdots & \pazocal{D}_{1,n}^T \\ \hline
\pazocal{D}_{0,2} & \pazocal{D}_{1,2} & \pazocal{L}_2 & \cdots & \pazocal{D}_{2,n}^T \\ \hline
\vdots & \vdots & \vdots & \ddots & \vdots \\ \hline
\pazocal{D}_{0,n} & \pazocal{D}_{1,n} & \pazocal{D}_{2,n} & \cdots & \pazocal{L}_n
\end{array}\right)
$$ 
where $\pazocal{D}_{p,r}=\sum_{i=0}^n D_{p,r}^i$ and $D_{p,r}^q(i,j)= s$, and $s$ is the number of the $q$-simplices that are adjacent to both $\sigma_j^p$ and $\sigma_i^r$ for $q \notin \{p,r\}$ and $D_{p,r}^p=D_{p,r}^r=D_{p,r}$.
Here, the blocks on the main diagonal are the Laplacians we develop initially, i.e., they provide the relation between simplices of fixed dimension through simplices of any dimension. Besides, the off-diagonal blocks do the same thing but for different dimensions. Therefore, $\pazocal{L}_H$ is able to capture the relations \textit{between} all simplices \textit{through} simplices of any dimension, which addresses all the limitations. 

To compute the influence of the higher-order interactions (i.e., hyperedges), we redefine four graph centrality measures, namely diffusion Frechet function (DFF) \cite{aktas2021influential, martinez2018probing}, degree, betweenness, and closeness~\cite{freeman1978centrality}, which are originally defined for vertices, to hyperedges, thanks to the generalized Laplacian $\pazocal{L}_H$. We use this Laplacian to model relations between hyperedges, and define the centrality measures accordingly.

DFF centrality employs the diffusion Fr{\'e}chet function (DFF) defined as the weighted sum of the diffusion distance between a hyperedge and the rest of the network. The diffusion distance measures the similarity between two given hyperedges by finding the similarity of the heat diffusion on a given time interval when the heat source is located on these hyperedges. More formally, let $\mathcal{E}=[\mathcal{E}_1,..., \mathcal{E}_n]^T \in \mathbb{R}^n$ be a probability distribution on hyperedge set $E$ of a hypergraph $H$. For $t>0$, the diffusion Fr{\'e}chet function on hyperedge $e_i \in E$ is defined as 
$$
F_{\mathcal{E},t}(i)=\sum_{j=1}^n d_t^2(i,j)\mathcal{E}_j.
$$
with 
$$
d_t^2(i,j)=\sum_{k=1}^n e^{-2\lambda_kt}(\phi_k(i) - \phi_k(j))^2
$$
where $0\leq \lambda_1 \leq ... \leq \lambda_n$ are the eigenvalues of the hypergraph Laplacian $\pazocal{L}_H$ with orthonormal eigenvectors $\phi_1,...,\phi_n$. A hyperedge with a smaller diffusion Fr{\'e}chet function value is considered an influential hyperedge in the network since the heat diffusion centered at this hyperedge is similar to many hyperedges. The degree centrality is the number of connections of each hyperedge. The degree can be computed considering the connection weights. The betweenness centrality measures how often each hyperedge appears on the shortest path between two hyperedges in the hypergraph. Since there can be several shortest paths between two hyperedges $s$ and $t$, the centrality of hyperedge $u$ is
    $$
   \displaystyle H_{Btw}(u)=\sum_{s,t \neq u} \frac{n_{st}(u)}{N_{st}}
    $$
where $n_{st}(u)$ is the number of shortest paths from $s$ to $t$ that pass through hyperedge $u$, and $N_{st}$ is the total number of shortest paths from $s$ to $t$. Shortest paths can be computed by considering the connection weights. Lastly, the closeness centrality uses the inverse sum of the distance from a hyperedge to all other hyperedges in the hypergraph. The centrality of a hyperedge $u$ is
    $$
    \displaystyle H_{Cls}(u)=\frac{1}{C_u}
    $$
where $C_u$ is the sum of the distances from hyperedge $u$ to all other hyperedges. The distance from a hyperedge to another hyperedge can be computed by considering the connection weights. 
 
Lastly, we model SIR on vertices of the hypergraph by utilizing the Laplacian $\pazocal{L}_{0}$ for evaluation. As we mentioned before, $\pazocal{L}_{0}$ reveals how vertices are connected through hyperedges of any size.

We now provide an illustrative example of the proposed methods.

\begin{figure}[h!]
    \centering
     \includegraphics[width=.4\textwidth]{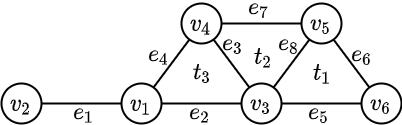}
    \caption{A hypergraph with six vertices (0-simplices), eight edges (1-simplices) and three triangles (2-simplex).}
    \label{fig:simpcomp}
\end{figure}
\begin{example}
The hypergraph in Figure \ref{fig:simpcomp} has six vertices (0-simplices), eight edges (1-simplices), and three triangles (2-simplices). Its Laplacian between 0-simplices, $\pazocal{L}_{0}$, and generalized Laplacian, $\pazocal{L}_{H}$, are found below. In both Laplacians, the diagonal entries show the number of neighboring simplices for each $k$-simplex (we also count each hyperedge as its neighboring hyperedge in order to stress the importance of the direct neighborhood relation), and the off-diagonal entries show the number of the shared neighboring simplices with other simplices. The diffusion between simplices happens based on the number of the shared neighboring simplices with other simplices in these Laplacians.

\[
\pazocal{L}_0=\begin{blockarray}{ccccccc}
& v_1 & v_2 & v_3 & v_4 & v_5 & v_6\\
\begin{block}{c(cccccc)} 
v_1 & 5 & 1 & 2 & 2 & 0 & 0 \\ 
v_2 & 1 & 2 & 0 & 0 & 0 & 0 \\ 
v_3 & 2 & 0 & 8 & 3 & 3 & 2 \\ 
v_4 & 2 & 0 & 3 & 6 & 2 & 0 \\ 
v_5 & 0 & 0 & 3 & 2 & 6 & 2 \\
v_6 & 0 & 0 & 2 & 0 & 2 & 4 \\
\end{block}
\end{blockarray}, \hspace{.5cm} \pazocal{L}_H=\left(
\begin{array}{cccccc|cccccccc|ccc}
5 & 1 & 2 & 2 & 0 & 0 & 2 & 3 & 1 & 3 & 0 & 0 & 0 & 0 & 4 & 0 & 0 \\
1 & 2 & 0 & 0 & 0 & 0 & 2 & 0 & 0 & 0 & 0 & 0 & 0 & 0 & 0 & 0 & 0 \\
2 & 0 & 8 & 3 & 3 & 2 & 0 & 3 & 4 & 1 & 3 & 1 & 1 & 4 & 4 & 4 & 4 \\
2 & 0 & 3 & 5 & 2 & 0 & 0 & 1 & 4 & 3 & 0 & 0 & 3 & 1 & 4 & 4 & 0 \\ 
0 & 0 & 3 & 2 & 5 & 2 & 0 & 0 & 1 & 0 & 1 & 3 & 3 & 4 & 0 & 4 & 4 \\
0 & 0 & 2 & 0 & 2 & 4 & 0 & 0 & 0 & 0 & 3 & 3 & 0 & 1 & 0 & 0 & 4 \\
\hline
2 & 2 & 0 & 0 & 0 & 0 & 3 & 1 & 0 & 1 & 0 & 0 & 0 & 0 & 1 & 0 & 0 \\
3 & 0 & 3 & 1 & 0 & 0 & 1 & 4 & 2 & 2 & 1 & 0 & 0 & 1 & 4 & 1 & 1 \\
1 & 0 & 4 & 4 & 1 & 0 & 0 & 2 & 5 & 2 & 1 & 0 & 2 & 2 & 4 & 4 & 1 \\
3 & 0 & 1 & 3 & 0 & 0 & 1 & 2 & 2 & 4 & 0 & 0 & 1 & 0 & 4 & 1 & 0 \\
0 & 0 & 3 & 0 & 1 & 3 & 0 & 1 & 1 & 0 & 4 & 2 & 0 & 2 & 1 & 1 & 4 \\
0 & 0 & 1 & 0 & 3 & 3 & 0 & 0 & 0 & 0 & 2 & 4 & 1 & 2 & 0 & 1 & 4 \\
0 & 0 & 1 & 3 & 3 & 0 & 0 & 0 & 2 & 1 & 0 & 1 & 4 & 2 & 1 & 4 & 1 \\
0 & 0 & 4 & 1 & 4 & 1 & 0 & 1 & 2 & 0 & 2 & 2 & 2 & 5 & 1 & 4 & 4 \\
\hline
4 & 0 & 4 & 4 & 0 & 0 & 1 & 4 & 4 & 4 & 1 & 0 & 1 & 1 & 7 & 3 & 1 \\
0 & 0 & 4 & 4 & 4 & 0 & 0 & 1 & 4 & 1 & 1 & 1 & 4 & 4 & 3 & 7 & 3 \\
0 & 0 & 4 & 0 & 4 & 4 & 0 & 1 & 1 & 0 & 4 & 4 & 1 & 4 & 1 & 3 & 7 \\
\end{array}\right).
\]

\begin{table}[h!]

\centering
\caption{The cells for the best result in each row is colored gray.}
\begin{tabular}{|c|c c|c c|c c|c c|c c|}

\hline   \multirow{2}{*}{$ $} & \multicolumn{2}{|c|}{\cellcolor{gray!60}$\mathbf{H}_{DFF}$} & \multicolumn{2}{|c|}{\cellcolor{gray!60}$\mathbf{H}_{Deg}$} & \multicolumn{2}{|c|}{\cellcolor{gray!60}$\mathbf{H}_{Btw}$} & \multicolumn{2}{|c|}{\cellcolor{gray!60}$\mathbf{H}_{Cls}$} & \multicolumn{2}{|c|}{\cellcolor{gray!60}$\mathbf{R}_s$} \\ \cline{2-11}
 & \cellcolor{gray!25}\textbf{Rank} & \cellcolor{gray!25}\textbf{Score}& \cellcolor{gray!25}\textbf{Rank} & \cellcolor{gray!25}\textbf{Score}& \cellcolor{gray!25}\textbf{Rank} & \cellcolor{gray!25}\textbf{Score}& \cellcolor{gray!25}\textbf{Rank} & \cellcolor{gray!25}\textbf{Score} & \cellcolor{gray!25}\textbf{Rank} & \cellcolor{gray!25}\textbf{Score} \\ \hline
\hline $e_1$ & 11 &	1.412 &	11 & 7 & 6 & 0.026 & 11	& 0.533	& 4	& 0.078	\\
\hline $e_2$ & 6 & 1.218 & 6 & 20 & 2 &	0.054 &	3 &	0.762 &	5 &	0.057 \\
\hline $e_3$ & 4 & 1.123 & 4 & 28 &	3 &	0.034 &	2 &	0.800 & 11 & 0.029 \\
\hline $e_4$ & 9 & 1.245 & 9 & 18 &	5 &	0.026 &	7 &	0.696 &	6 &	0.052 \\
\hline $e_5$ & 8 & 1.231 & 8 & 19 &	9 &	0.017 &	8 &	0.696 &	7 &	0.045 \\
\hline $e_6$ & 10	&	1.259	&	10	&	17	&	11	&	0.004	&	10	&	0.615	&	8	&	0.044	\\
\hline $e_7$ & 7	&	1.231	&	7	&	19	&	10	&	0.015	&	9	&	0.696	&	9	&	0.040	\\
\hline $e_8$ & 5	&	1.123	&	5	&	28	&	4	&	0.028	&	4	&	0.762	&	10	&	0.031	\\
\hline $t_1$ & 3	&	1.092	&	3	&	31	&	8	&	0.021	&	6	&	0.727	&	2	&	0.125	\\
\hline $t_2$ & 1	&	1.061	&	1	&	34	&	7	&	0.024	&	5	&	0.762	&	3	&	0.111	\\
\hline $t_3$ & 2	&	1.081	&	2	&	32	&	1	&	0.065	&	1	&	0.800	&	1	&	0.139	\\

\hline 
\end{tabular}
\label{table:toy}
\end{table}

To find the influential higher-order interactions, we apply the SIR model and calculate centralities using $\pazocal{L}_0$ and $\pazocal{L}_H$, respectively. As we see in Table \ref{table:toy}, $t_1, t_2, t_3$ are the most influential higher-order interactions based on $H_{DFF}$ and $H_{Deg}$ and these results are aligned with the corresponding diffusion index $R_s$. Similarly, $t_3$ is the most influential based on $H_{Btw}$ and $H_{Cls}$. On the other hand, although $e_3$ can be considered as one of the most central and, as a result, influential higher-order interaction, its diffusion index is relatively low. The reason for this is that its neighboring simplices, $v_3, v_4, t_2, t_3$, are of high degree; hence, its removal relatively affects the diffusion. 
\end{example}

\section{Results}
In this section, we first describe the datasets we use in our experiments. Then, we explain how we model SIR on hypergraphs using the proposed Laplacians for evaluation. Next, we redefine four classical graph centrality measures, namely DFF, degree, betweenness, and closeness, to hypergraphs and present results on the datasets.  

\noindent\textbf{Data Description.} In our experiments, we use four undirected real-world networks to evaluate the effectiveness of redefined centrality measures using hypergraph Laplacians (see Table~\ref{table:data} for the networks' statistics). (1) Enron: each vertex represents the email address of a staff member at Enron. A simplex or hyperedge represents all the recipients, including the sender, of an email sent between the Enron staff. (2) High school: this dataset is made from a network of high school students in Marseilles, France. A vertex is a student, and a simplex is a set of students in close contact with each other. (3) Primary school: this dataset is made from a network of primary school students and teachers. A vertex is a student or a teacher, and a simplex is a set of students and/or teachers in close contact with each other. (4) NDC-classes: a vertex is a pharmaceutical class label used to classify a certain property of a drug. The network of drugs is taken from the National Drug Code Directory. A simplex is a set of several or many class labels assigned to a drug. These datasets can be found in~\cite{benson2018simplicial}. 

\begin{table}[h!]
\centering
\caption{Basic properties of the real-world datasets we use are provided here. $\langle k\rangle$ is the average weighted degree, and $k_{max}$ is the maximum hyperedge size.}

\begin{tabular}{|c|c|c|c|c|}
\hline  $Dataset$ & $Vertices$ & $Hyperedges$ & $\langle k\rangle$ & $k_{max}$ \\ \hline
\hline Enron & 143 & 1630 & 106.3 & 19 \\
\hline High school & 327 & 8264 & 81.9 & 6 \\
\hline Primary school & 242 & 13041 & 197.2 & 6 \\
\hline NDC-classes & 1149 & 2330 & 62.5 & 25 \\
\hline
\end{tabular}

\label{table:data}
\end{table}

\noindent\textbf{Evaluation metrics.} We use the Susceptible-Infected-Recovered (SIR) simulation model as an evaluation metric to objectively analyze the effect of higher-order interactions in diffusion between nodes. In the SIR model, each node is classified as a Susceptible (S), Infected (I), or Recovered (R) at any given moment. A selected node is initially infected, and the rest of the network is susceptible to be infected. In each propagation, the infected node can infect its neighboring nodes with probability $\mu$. As this process is repeated, infected nodes can recover with probability $\beta$ and are not susceptible to be infected again. The number of nodes that were infected in the network measures the diffusion level. A greater number means a greater spreading ability and a greater influence on diffusion. 

In our experiments with the SIR models, we set infection rate based on $\mu_c$, where $\mu_c=\frac{\langle k\rangle}{\langle k^2\rangle-\langle k\rangle}$, as derived from \cite{guo2020influential}, and $\langle k\rangle$ is equal to the average weighted degree of the network. Furthermore, for the infection rate, we can also consider the weights of interactions. In this paper, we set the infection rate of an interaction of weight $w > 0$ to $\mu_w=1-(1-\mu)^w$, following \cite{sun2014epidemic}. For simplicity, the recovery rate is set as $\beta =1$. The experiment is run 100 times for each dataset, and the average of the 100 trials is taken to obtain more reliable results. 

As \cite{yu2018identifying} suggests, we use the normalized final effected scale for evaluation, which is defined as
\[
\displaystyle F(u)=\frac{n_u}{n}
\]
where $n_u$ is the number of affected nodes when node $u$ is infected, and $n$ is the total number of nodes. To compute the influence of higher-order interactions, we calculate the average influence of all nodes after removing a certain fraction of hyperedges as the following diffusion index
\[
\displaystyle R_s=\frac{F_1 - F_2}{F_1}
\]
where $F_i$ is the average final infected scale of all nodes, i.e., $F_i=\frac{1}{n}\sum_{u \in V} F(u)$ for $i \in \{1,2\}$, and $F_1$ and $F_2$ are results of the original network and the network after removing $p$ of higher-order interactions respectively.

In our experiment, we obtain three different results. For the first result, we first rank the higher-order interactions from the most influential to the least based on the centralities and divide them into 50 parts. For each step, we only remove one part and calculate the diffusion index. Finally, we sort the diffusion indices of the 50 parts for each centrality and find its spearman correlation coefficients to the centrality scores. As we see in Table \ref{table:spea}, the proposed hypergraph centrality measures can find the influential higher-order interactions in diffusion effectively. All the correlations are high for the Enron dataset and are about 90\%, except for 85\% for $H_{DFF}$. In the High school dataset, all of the centrality measures' rankings are highly correlated with the SIR findings: they are all about $98 \%$. It is the same for the Primary school dataset, except $H_{Btw}$. The correlations are a little lower for the NDC-classes dataset than the other datasets (it is in between \%76-\%80 for all but about \%55 for $H_{Btw}$). The reason is that since the ratio between the number of nodes and simplices is low in this dataset, the infection rate becomes relatively small. This makes diffusion difficult in the SIR simulations, and as a result, it makes it slightly more difficult to see the effects of removing higher-order interactions. In general, $H_{DFF}$, $H_{Deg}$, and $H_{Cls}$ provide similar effectiveness, and $H_{Btw}$ is slightly less effective on the NDC-classes and Primary school datasets. We should also note here that our main goal here is not to compare these proposed hypergraph centralities but to show how effective they are in finding influential higher-order interactions. As we see in Table \ref{table:spea}, overall, they are quite effective in finding influential higher-order interactions.

\begin{table}[h!]
\centering
\caption{Spearman correlation coefficients between the ranking scores and the diffusion indices. The results are averaged over 100 independent implementations with $\mu/\mu_c=1.5$.}
\begin{tabular}{|c|c|c|c|c|}

\hline   & $H_{DFF}$ & $H_{Deg}$ & $H_{Btw}$ & $H_{Cls}$ \\ \hline
\hline Enron & \%85.87 & \%90.25 & \%90.67 & \%90.55  \\
\hline High school & \%98.93 & \%97.79 & \%98.39 & \%98.54    \\
\hline Primary school & \%97.64 & \%98.05 & \%83.64  & \%95.58  \\
\hline NDC-classes & \%79.43 & \%76.84 & \%54.32 & \% 78.11 \\

\hline
\end{tabular}
\label{table:spea}
\end{table}

\begin{figure}[h!]
    \centering
     \includegraphics[width=\textwidth]{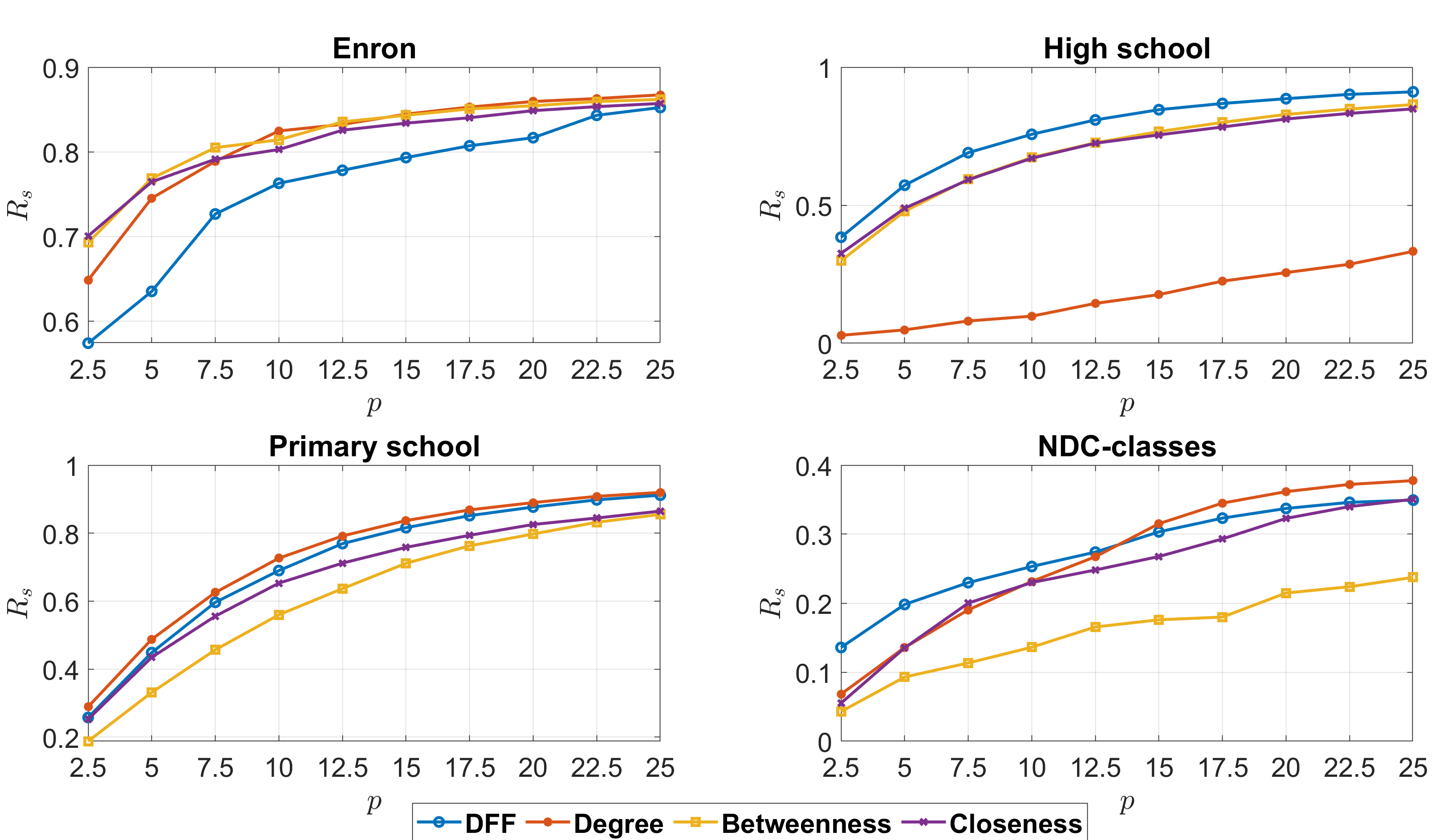}
    \caption{Varying ratio of simplices through the implementation of the SIR Model. Here the infection rate is kept constant while the influential simplices are obtained by using each of the centrality measures. A higher diffusion index ($R_s$) determines the effectiveness of each of these methods.}
    \label{fig:ratio}
\end{figure}

For the second result, we fix the infection rate at $\mu=1.5\mu_c$ and vary the ratio of the removed influential higher-order interactions in Figure~\ref{fig:ratio}. The graphs are plotted with the diffusion index $R_s$ on the $y$-axes and the varying ratio of simplices ($p$) on the $x$-axes. When analyzing the graph for the Enron dataset, it can be inferred that degree, betweenness, and closeness are seen to overlap each other and possesses a higher diffusion index $R_s$ in comparison to DFF centrality. Therefore, it can be concluded that for the Enron dataset, degree, betweenness, and closeness centralities can be considered to be almost equally effective due to a consistently high $R_s$ value with the increasing ratio of simplices. For the High School dataset, it can be observed that DFF centrality is the most effective of the four centrality measures with betweeness and closeness centralities overlapping and following the same set of values for $R_s$ with an increase in the varying ratios. In contrast, the degree centrality proved to be least effective for its low $R_s$ value below 0.5. Primary school dataset has the most consistent increase in the value of diffusion index $R_s$ for all the centrality measures as there are little or no anomalies within the curves plotted for the four different centrality measures. The degree and the DFF centralities are considered the most effective centralities as they both yield the highest value of $R_s$ in comparison to betweenness and closeness centralities. Finally, for the NDC-classes dataset, it can be seen that initially the DFF centrality had a higher diffusion index ($R_s$) in comparison to the degree centrality. Still, after the ratio ($p$) increases above 12.5, the degree centrality proves to be more effective than the DFF centrality and the other two centrality measures. In conclusion, all the four centrality measures proved to be effective with varying ratios of higher-order interactions used within this SIR model.

\begin{figure}[h!]
    \centering
     \includegraphics[width=\textwidth]{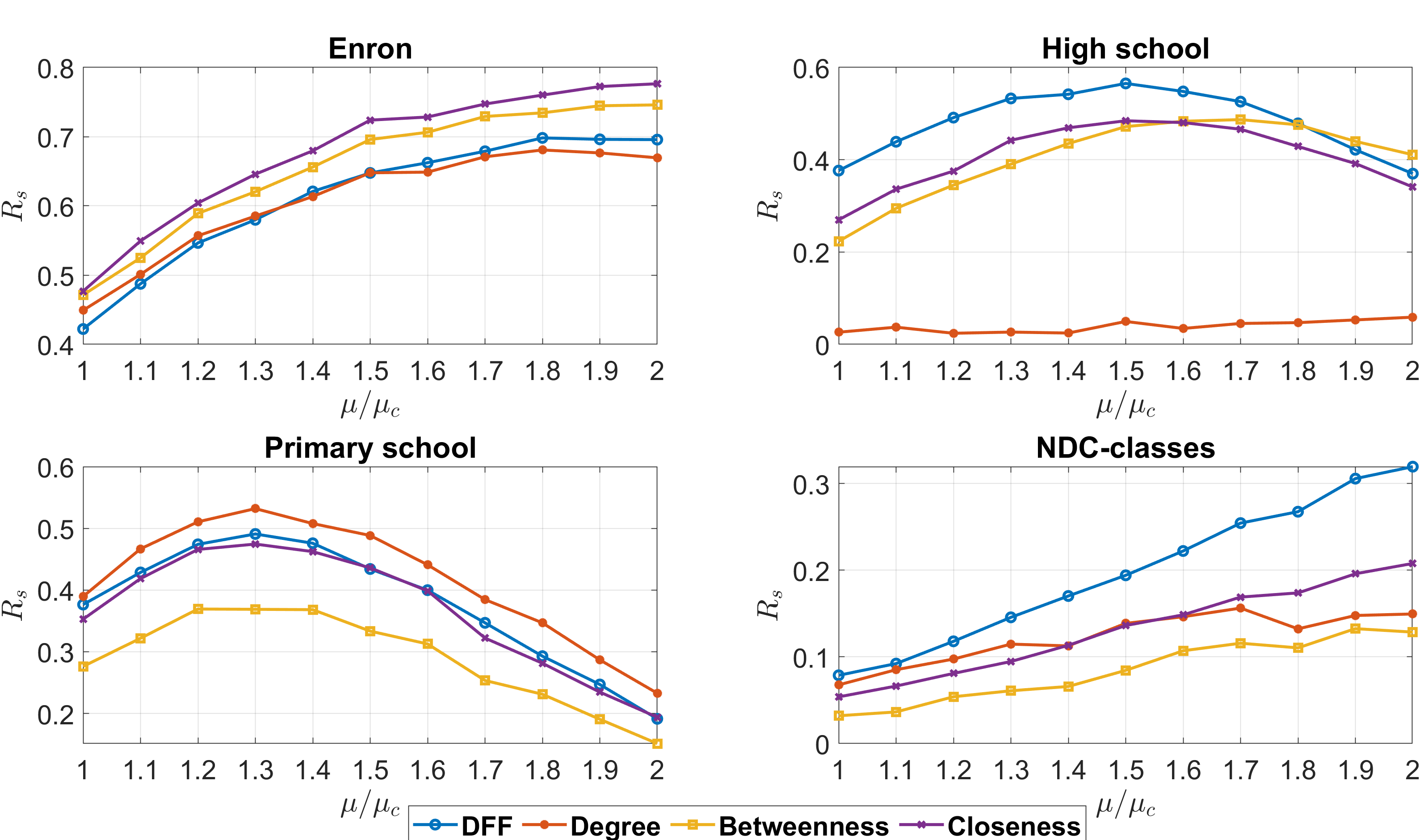}
    \caption{Varying infection rate when using the SIR model on the selected networks. The ratio of simplices removed in each trial is the top 5\% of influential simplices found by each of the centrality measures. A greater diffusion index or $R_s$ indicates that the method is more effective. }
    \label{fig:inf}
\end{figure}

For the third result, we fix the ratio of the removed higher-order interactions at $5\%$ and vary the infection rate to compare the centrality measures further in Figure \ref{fig:inf}. A greater diffusion index or $R_s$ indicates that the removed simplices are influential, and therefore, the method is more effective. With the Enron dataset, closeness centrality performs the most effectively with betweenness centrality after it. DFF overall performs about as well as degree centrality but is outperformed by betweenness centrality and closeness centrality. However, the differences in performances are not drastic, and all the centrality measures overall perform well. In the Primary school network, degree centrality outperforms all the other centrality measures. Closeness and DFF overlap at several infection rates, with DFF performing slightly more effectively at other infection rates. In the High school dataset, DFF outperforms all the other centrality measures. However, after the infection rate reaches 1.8, betweenness centrality performs better than the rest of the centrality measures. Closeness centrality also performs well. In this dataset, degree centrality does not perform as well. When analyzing the NDC-classes network, it can be concluded that DFF  outperforms all the centrality measures. Degree centrality performs better than closeness centrality at the lower infection rates. The two centrality measures' effectiveness is about the same when the infection rate is set in between 1.4-1.6. At the greater infection rates, closeness outperforms degree centrality. Betweenness centrality is the least effective of the centrality measures in this dataset, but performance is not drastic. It can be noted that with Enron and NDC-classes, as the infection rate increases, the diffusion index or $R_s$ increases as well. However, for High school and Primary school, $R_s$ initially increases but then decreases. This is due to the network structure of these two datasets. Overall, the centrality methods, with varying performances depending on the network applied to, effectively identify influential higher-order interactions in complex networks for various infection rates used in the SIR model. 

\section{Discussion}

By proposing two new hypergraph Laplacians, we are able to redefine DFF, betweenness centrality, closeness centrality, and degree centrality to determine the influential higher-order interactions of a network. These centrality measures are applied to four real-world network datasets. The performances of the centrality measures in identifying influential higher-order network interactions are compared and evaluated by the SIR model and using spearman's rank correlation coefficients. Overall, all the centrality measures, adjusted to work with higher-order simplices, are effective in finding influential higher-order interactions. The high spearman correlations values for the centrality measures indicate this effectiveness as in Table \ref{table:spea}. Furthermore, the methods are effective when varying the ratio of infected simplicies and when varying infection rates.

As mentioned earlier, there are not many known centrality measures capable of effectively analyzing and identifying influential higher-order interactions in complex networks. Being able to utilize several centrality measures provides more flexibility, expanding the uses of determining significant higher-order interactions. It also provides more options to select the best possible method of analyzing higher-order interactions based on specific types of networks as well as computational complexity and more room to expand on methods for finding influential higher-order interactions. The results found are significant as they provide a basis for DFF, betweenness centrality, closeness centrality, and degree centrality, through hypergraph Laplacians, to be utilized in real-life applications such as rumor controlling, marketing, disease spreading, advertising, and more.

\bibliographystyle{unsrtnat}
\bibliography{references}  

\begin{thebibliography}{32}
\providecommand{\natexlab}[1]{#1}
\providecommand{\url}[1]{\texttt{#1}}
\expandafter\ifx\csname urlstyle\endcsname\relax
  \providecommand{\doi}[1]{doi: #1}\else
  \providecommand{\doi}{doi: \begingroup \urlstyle{rm}\Url}\fi

\bibitem[Kempe et~al.(2003)Kempe, Kleinberg, and Tardos]{kempe2003maximizing}
David Kempe, Jon Kleinberg, and {\'E}va Tardos.
\newblock Maximizing the spread of influence through a social network.
\newblock In \emph{Proceedings of the ninth ACM SIGKDD international conference
  on Knowledge discovery and data mining}, pages 137--146, 2003.

\bibitem[Wang et~al.(2020)Wang, Gong, Liu, and Wu]{wang2020preventing}
Shanfeng Wang, Maoguo Gong, Wenfeng Liu, and Yue Wu.
\newblock Preventing epidemic spreading in networks by community detection and
  memetic algorithm.
\newblock \emph{Applied Soft Computing}, 89:\penalty0 106118, 2020.

\bibitem[Bonacich(1972)]{bonacich1972factoring}
Phillip Bonacich.
\newblock Factoring and weighting approaches to status scores and clique
  identification.
\newblock \emph{Journal of mathematical sociology}, 2\penalty0 (1):\penalty0
  113--120, 1972.

\bibitem[L{\"u} et~al.(2016)L{\"u}, Zhou, Zhang, and Stanley]{lu2016h}
Linyuan L{\"u}, Tao Zhou, Qian-Ming Zhang, and H~Eugene Stanley.
\newblock The h-index of a network node and its relation to degree and
  coreness.
\newblock \emph{Nature communications}, 7\penalty0 (1):\penalty0 1--7, 2016.

\bibitem[Freeman(1978)]{freeman1978centrality}
Linton~C Freeman.
\newblock Centrality in social networks conceptual clarification.
\newblock \emph{Social networks}, 1\penalty0 (3):\penalty0 215--239, 1978.

\bibitem[Freeman(1977)]{freeman1977set}
Linton~C Freeman.
\newblock A set of measures of centrality based on betweenness.
\newblock \emph{Sociometry}, pages 35--41, 1977.

\bibitem[Brin and Page(1998)]{brin1998anatomy}
Sergey Brin and Lawrence Page.
\newblock The anatomy of a large-scale hypertextual web search engine.
\newblock \emph{Computer networks and ISDN systems}, 30\penalty0
  (1-7):\penalty0 107--117, 1998.

\bibitem[Aktas et~al.()Aktas, Jawaid, Harrington, and
  Akbas]{aktas2021influential}
Mehmet~E Aktas, Sidra Jawaid, Ebony Harrington, and Esra Akbas.
\newblock Influential nodes detection in complex networks via diffusion
  fr{\'e}chet function.
\newblock under review.

\bibitem[Arulselvan et~al.(2009)Arulselvan, Commander, Elefteriadou, and
  Pardalos]{arulselvan2009detecting}
Ashwin Arulselvan, Clayton~W Commander, Lily Elefteriadou, and Panos~M
  Pardalos.
\newblock Detecting critical nodes in sparse graphs.
\newblock \emph{Computers \& Operations Research}, 36\penalty0 (7):\penalty0
  2193--2200, 2009.

\bibitem[Yang et~al.(2019)Yang, Benko, Cavaliere, Huang, and
  Perc]{yang2019identification}
Guoli Yang, Tina~P Benko, Matteo Cavaliere, Jincai Huang, and Matja{\v{z}}
  Perc.
\newblock Identification of influential invaders in evolutionary populations.
\newblock \emph{Scientific reports}, 9\penalty0 (1):\penalty0 1--12, 2019.

\bibitem[Giuraniuc et~al.(2005)Giuraniuc, Hatchett, Indekeu, Leone, Castillo,
  Van~Schaeybroeck, and Vanderzande]{giuraniuc2005trading}
CV~Giuraniuc, JPL Hatchett, JO~Indekeu, M~Leone, I~P{\'e}rez Castillo, Bert
  Van~Schaeybroeck, and Carlo Vanderzande.
\newblock Trading interactions for topology in scale-free networks.
\newblock \emph{Physical review letters}, 95\penalty0 (9):\penalty0 098701,
  2005.

\bibitem[Girvan and Newman(2002)]{girvan2002community}
Michelle Girvan and Mark~EJ Newman.
\newblock Community structure in social and biological networks.
\newblock \emph{Proceedings of the national academy of sciences}, 99\penalty0
  (12):\penalty0 7821--7826, 2002.

\bibitem[Wang et~al.(2017)Wang, He, Nechifor, Zhang, and
  Crossley]{wang2017identification}
Ziqi Wang, Jinghan He, Alexandru Nechifor, Dahai Zhang, and Peter Crossley.
\newblock Identification of critical transmission lines in complex power
  networks.
\newblock \emph{Energies}, 10\penalty0 (9):\penalty0 1294, 2017.

\bibitem[Zio et~al.(2012)Zio, Golea, et~al.]{zio2012identifying}
Enrico Zio, Lucia~R Golea, et~al.
\newblock Identifying groups of critical edges in a realistic electrical
  network by multi-objective genetic algorithms.
\newblock \emph{Reliability Engineering \& System Safety}, 99:\penalty0
  172--177, 2012.

\bibitem[Saito et~al.(2016)Saito, Kimura, Ohara, and
  Motoda]{saito2016detecting}
Kazumi Saito, Masahiro Kimura, Kouzou Ohara, and Hiroshi Motoda.
\newblock Detecting critical links in complex network to maintain information
  flow/reachability.
\newblock In \emph{Pacific Rim International Conference on Artificial
  Intelligence}, pages 419--432. Springer, 2016.

\bibitem[Wong et~al.(2017)Wong, Sun, Lo, Yiu, Wu, Zhao, Chan, and
  Kao]{wong2017finding}
Petrie Wong, Cliz Sun, Eric Lo, Man~Lung Yiu, Xiaowei Wu, Zhichao Zhao,
  T-H~Hubert Chan, and Ben Kao.
\newblock Finding k most influential edges on flow graphs.
\newblock \emph{Information Systems}, 65:\penalty0 93--105, 2017.

\bibitem[Hamers et~al.(1989)]{hamers1989similarity}
Lieve Hamers et~al.
\newblock Similarity measures in scientometric research: The jaccard index
  versus salton's cosine formula.
\newblock \emph{Information Processing and Management}, 25\penalty0
  (3):\penalty0 315--18, 1989.

\bibitem[Yu et~al.(2018)Yu, Chen, and Zhao]{yu2018identifying}
En-Yu Yu, Duan-Bing Chen, and Jun-Yan Zhao.
\newblock Identifying critical edges in complex networks.
\newblock \emph{Scientific reports}, 8\penalty0 (1):\penalty0 1--8, 2018.

\bibitem[Cheng et~al.(2010)Cheng, Ren, Shen, Zhang, and
  Zhou]{cheng2010bridgeness}
Xue-Qi Cheng, Fu-Xin Ren, Hua-Wei Shen, Zi-Ke Zhang, and Tao Zhou.
\newblock Bridgeness: a local index on edge significance in maintaining global
  connectivity.
\newblock \emph{Journal of Statistical Mechanics: Theory and Experiment},
  2010\penalty0 (10):\penalty0 P10011, 2010.

\bibitem[Battiston et~al.(2020)Battiston, Cencetti, Iacopini, Latora, Lucas,
  Patania, Young, and Petri]{battiston2020networks}
Federico Battiston, Giulia Cencetti, Iacopo Iacopini, Vito Latora, Maxime
  Lucas, Alice Patania, Jean-Gabriel Young, and Giovanni Petri.
\newblock Networks beyond pairwise interactions: structure and dynamics.
\newblock \emph{arXiv preprint arXiv:2006.01764}, 2020.

\bibitem[Veremyev et~al.(2017)Veremyev, Prokopyev, and
  Pasiliao]{veremyev2017finding}
Alexander Veremyev, Oleg~A Prokopyev, and Eduardo~L Pasiliao.
\newblock Finding groups with maximum betweenness centrality.
\newblock \emph{Optimization Methods and Software}, 32\penalty0 (2):\penalty0
  369--399, 2017.

\bibitem[Veremyev et~al.(2019)Veremyev, Prokopyev, and
  Pasiliao]{veremyev2019finding}
Alexander Veremyev, Oleg~A Prokopyev, and Eduardo~L Pasiliao.
\newblock Finding critical links for closeness centrality.
\newblock \emph{INFORMS Journal on Computing}, 31\penalty0 (2):\penalty0
  367--389, 2019.

\bibitem[Vogiatzis et~al.(2015)Vogiatzis, Veremyev, Pasiliao, and
  Pardalos]{vogiatzis2015integer}
Chrysafis Vogiatzis, Alexander Veremyev, Eduardo~L Pasiliao, and Panos~M
  Pardalos.
\newblock An integer programming approach for finding the most and the least
  central cliques.
\newblock \emph{Optimization Letters}, 9\penalty0 (4):\penalty0 615--633, 2015.

\bibitem[Nasirian et~al.(2020)Nasirian, Pajouh, and
  Balasundaram]{nasirian2020detecting}
Farzaneh Nasirian, Foad~Mahdavi Pajouh, and Balabhaskar Balasundaram.
\newblock Detecting a most closeness-central clique in complex networks.
\newblock \emph{European Journal of Operational Research}, 283\penalty0
  (2):\penalty0 461--475, 2020.

\bibitem[Zhao et~al.(2017)Zhao, Wang, Lui, Towsley, and
  Guan]{zhao2017efficient}
Junzhou Zhao, Pinghui Wang, John~CS Lui, Don Towsley, and Xiaohong Guan.
\newblock I/o-efficient calculation of h-group closeness centrality over
  disk-resident graphs.
\newblock \emph{Information Sciences}, 400:\penalty0 105--128, 2017.

\bibitem[Newman(2002)]{newman2002spread}
Mark~EJ Newman.
\newblock Spread of epidemic disease on networks.
\newblock \emph{Physical review E}, 66\penalty0 (1):\penalty0 016128, 2002.

\bibitem[Horak and Jost(2013)]{horak2013spectra}
Danijela Horak and J{\"u}rgen Jost.
\newblock Spectra of combinatorial laplace operators on simplicial complexes.
\newblock \emph{Advances in Mathematics}, 244:\penalty0 303--336, 2013.

\bibitem[Aktas and Akbas(2021)]{aktas2021hypergraph}
Mehmet~Emin Aktas and Esra Akbas.
\newblock Hypergraph laplacians in diffusion framework.
\newblock \emph{arXiv preprint arXiv:2102.08867}, 2021.

\bibitem[Mart{\'\i}nez et~al.(2018)Mart{\'\i}nez, Lee, Kim, and
  Mio]{martinez2018probing}
Diego H~D{\'\i}az Mart{\'\i}nez, Christine~H Lee, Peter~T Kim, and Washington
  Mio.
\newblock Probing the geometry of data with diffusion fr{\'e}chet functions.
\newblock \emph{Applied and Computational Harmonic Analysis}, 2018.
\newblock Elsevier.

\bibitem[Benson et~al.(2018)Benson, Abebe, Schaub, Jadbabaie, and
  Kleinberg]{benson2018simplicial}
Austin~R Benson, Rediet Abebe, Michael~T Schaub, Ali Jadbabaie, and Jon
  Kleinberg.
\newblock Simplicial closure and higher-order link prediction.
\newblock \emph{Proceedings of the National Academy of Sciences}, 115\penalty0
  (48):\penalty0 E11221--E11230, 2018.

\bibitem[Guo et~al.(2020)Guo, Yang, Chen, Chen, Gao, and
  Ma]{guo2020influential}
Chungu Guo, Liangwei Yang, Xiao Chen, Duanbing Chen, Hui Gao, and Jing Ma.
\newblock Influential nodes identification in complex networks via information
  entropy.
\newblock \emph{Entropy}, 22\penalty0 (2):\penalty0 242, 2020.

\bibitem[Sun et~al.(2014)Sun, Liu, Zhang, and Zhang]{sun2014epidemic}
Ye~Sun, Chuang Liu, Chu-Xu Zhang, and Zi-Ke Zhang.
\newblock Epidemic spreading on weighted complex networks.
\newblock \emph{Physics Letters A}, 378\penalty0 (7-8):\penalty0 635--640,
  2014.

\end{thebibliography}






\end{document}